\documentclass[11pt]{article}

\usepackage{graphicx}
\usepackage{cite}
\usepackage{mathpple}

\usepackage{geometry}
\begin{document}

{\small\noindent
\textbf{Preprint of:}\\
T. A. Nieminen, H. Rubinsztein-Dunlop and N. R. Heckenberg\\
``Calculation of the \textit{T}-matrix: general considerations and
application of the point-matching method''\\
\textit{Journal of Quantitative Spectroscopy and Radiative Transfer}
\textbf{79--80}, 1019--1029 (2003)
}

\vspace{6mm}

\hrulefill

\vspace{9mm}

\begin{center}

\LARGE
\textbf{Calculation of the \textit{T}-matrix: general considerations and
application of the point-matching method}

\vspace{3mm}

\large
T. A. Nieminen, H. Rubinsztein-Dunlop, and N. R. Heckenberg

\vspace{3mm}

\normalsize
\textit{Centre for Biophotonics and Laser Science, Department of Physics,\\
The University of Queensland, Brisbane QLD 4072, Australia}

\texttt{timo@physics.uq.edu.au}

\end{center}

\begin{abstract}
The \textit{T}-matrix method is widely used for the calculation of
scattering by particles of sizes on the order of the illuminating wavelength.
Although the extended boundary condition method (EBCM) is the most commonly used
technique for calculating the \textit{T}-matrix, a variety of methods can be
used.

We consider some general principles of calculating \textit{T}-matrices, and
apply the point-matching method to calculate the \textit{T}-matrix for
particles devoid of symmetry. This method avoids the time-consuming
surface integrals required by the EBCM.

\vspace{3mm}


\end{abstract}

\section{The \textit{T}-matrix method}

The \textit{T}-matrix method in wave scattering involves writing the
relationship between the wave incident upon a scatterer, expanded in terms of
orthogonal eigenfunctions,
\begin{equation}
U_\mathrm{inc} = \sum_n^\infty a_n \psi_n^{(\mathrm{inc})},
\end{equation}
where $a_n$ are the expansion coefficients for the incident wave,
and the scattered wave, also expanded in terms of orthogonal eigenfunctions,
\begin{equation}
U_\mathrm{scat} = \sum_k^\infty p_k \psi_k^{(\mathrm{scat})},
\end{equation}
where $p_k$ are the expansion coefficients for the scattered wave,
is written as a simple matrix equation
\begin{equation}
p_k = \sum_n^\infty T_{kn} a_n
\end{equation}
or,  in more concise notation,
\begin{equation}
\mathbf{P} = \mathbf{T} \mathbf{A}
\end{equation}
where $T_{kn}$ are the elements of the \textit{T}-matrix.
The \textit{T}-matrix method can be used for scalar waves or vector waves in a
variety of geometries, with the only restrictions being that the geometry of
the problem permits expansion of the waves as discrete series in terms of
orthogonal eigenfunctions, that the response of the scatterer to the incident
wave is linear, and that the expansion series for the waves can be truncated at
a finite number of terms. In general, one calculates the \textit{T}-matrix,
although it is conceivable that it might be measured experimentally.

The \textit{T}-matrix depends only on the particle---its composition,  size,
shape, and orientation---and is independent of the incident field. This means
that for any particular particle, the \textit{T}-matrix only needs to be
calculated once, and can then be used for repeated calculations. This is a
significant advantage over many other methods of calculating scattering where
the entire calculation needs to be repeated~\cite{mishchenko_book}. Some cases
provide even more efficiency: if the waves are expanded in spherical functions,
the averaging of scattering over various orientations of the particle compared
to the direction of the incident wave can be performed
analytically~\cite{mishchenko1991}.

In the spherical geometry of elastic light scattering by a particle contained
entirely within some radius $r_0$, the eigenfunction expansions of the fields
are made in terms of vector spherical wavefunctions
(VSWFs)~\cite{mishchenko_book,%
mishchenko1991,waterman1971,tsang1985,jackson,varshalovich,brock}:
\begin{eqnarray} \mathbf{M}_{nm}^{(1,2)}(k\mathrm{r}) & = & N_n h_n^{(1,2)}(kr)
\mathbf{C}_{nm}(\theta,\phi) \\
\mathbf{N}_{nm}^{(1,2)}(k\mathrm{r}) & = & \frac{h_n^{(1,2)}(kr)}{krN_n}
\mathbf{P}_{nm}(\theta,\phi) + \nonumber \\
& & N_n \left( h_{n-1}^{(1,2)}(kr) -
\frac{n h_n^{(1,2)}(kr)}{kr} \right) \mathbf{B}_{nm}(\theta,\phi)
\end{eqnarray}
where $h_n^{(1,2)}(kr)$ are spherical Hankel functions of the first and
second kind, $N_n = 1/\sqrt{n(n+1)}$ are normalisation constants, and
$\mathbf{B}_{nm}(\theta,\phi) = \mathbf{r} \nabla Y_n^m(\theta,\phi)$,
$\mathbf{C}_{nm}(\theta,\phi) = \nabla \times \left( \mathbf{r}
Y_n^m(\theta,\phi) \right)$, and
$\mathbf{P}_{nm}(\theta,\phi) = \hat{\mathbf{r}} Y_n^m(\theta,\phi)$
are the vector spherical harmonics~\cite{mishchenko_book,%
mishchenko1991,waterman1971,tsang1985,jackson,varshalovich,brock},
and $Y_n^m(\theta,\phi)$ are normalised scalar spherical harmonics. The usual
polar spherical coordinates are used, where $\theta$ is the co-latitude measured
from the $+z$ axis, and $\phi$ is the azimuth, measured from the $+x$ axis
towards the $+y$ axis.

$\mathbf{M}_{nm}^{(1)}$ and $\mathbf{N}_{nm}^{(1)}$ are outward-propagating TE
and TM multipole fields, while $\mathbf{M}_{nm}^{(2)}$ and
$\mathbf{N}_{nm}^{(2)}$ are the corresponding inward-propagating multipole
fields. Since these wavefunctions are purely incoming and purely outgoing, each
has a singularity at the origin. Since fields that are free of singularities
are of interest, it is useful to define the
singularity-free regular vector spherical wavefunctions:
\begin{eqnarray}
\mathbf{RgM}_{nm}(k\mathrm{r}) & = & {\textstyle\frac{1}{2}} [
\mathbf{M}_{nm}^{(1)}(k\mathrm{r}) + \mathbf{M}_{nm}^{(2)}(k\mathrm{r}) ], \\
\mathbf{RgN}_{nm}(k\mathrm{r}) & = & {\textstyle\frac{1}{2}} [
\mathbf{N}_{nm}^{(1)}(k\mathrm{r}) + \mathbf{N}_{nm}^{(2)}(k\mathrm{r}) ].
\end{eqnarray}
Since the spherical Bessel functions $j_n(kr) = \frac{1}{2}(h_n^{(1)}(kr) +
h_n^{(2)}(kr))$, the regular VSWFs are identical to the incoming and outgoing
VSWFs except for the replacement of the spherical Hankel functions by spherical
Bessel functions.

Since the incident field, in the absence of a scatterer, is singularity-free,
the expansion
\begin{equation}
\mathbf{E}_\mathrm{inc}(\mathrm{r}) = \sum_{n=1}^\infty \sum_{m = -n}^n
a^{(3)}_{nm} \mathbf{RgM}_{nm}(k\mathrm{r}) +
b^{(3)}_{nm} \mathbf{RgN}_{nm}(k\mathrm{r})
\label{regular_expansion}
\end{equation}
is generally used for the incident field.
Alternatively, the purely incoming part of the incident field can be used:
\begin{equation}
\mathbf{E}_\mathrm{inc}(\mathrm{r}) = \sum_{n=1}^\infty \sum_{m = -n}^n
a^{(2)}_{nm} \mathbf{M}_{nm}^{(2)}(k\mathrm{r}) +
b^{(2)}_{nm} \mathbf{N}_{nm}^{(2)}(k\mathrm{r}).
\label{incoming_expansion}
\end{equation}
In both cases, the scattered field is
\begin{equation}
\mathbf{E}_\mathrm{scat}(\mathrm{r}) = \sum_{n=1}^\infty \sum_{m = -n}^n
p^{(1)}_{nm} \mathbf{M}_{nm}^{(1)}(k\mathrm{r}) +
q^{(1)}_{nm} \mathbf{N}_{nm}^{(1)}(k\mathrm{r}).
\label{outgoing_expansion}
\end{equation}
The two sets of expansion coefficients for the incident/incoming field are
related, since $a^{(3)}_{nm} = 2a^{(2)}_{nm}$ and $b^{(3)}_{nm} =
2b^{(2)}_{nm}$. However, the scattered/outgoing field expansion coefficients
will differ, as will the \textit{T}-matrix. Using the regular expansion, the
\textit{T}-matrix in the absence of a scatterer is a zero matrix, while using
the incoming field expansion, the no-scatterer \textit{T}-matrix is the identity
matrix. The two expansions are essentially the same---the only difference is
that the incident wave in the incident/scattered wave expansion includes
part of the outgoing wave. \textit{T}-matrices for the two expansions only
differ by the identity matrix, so $\mathbf{T}^{\mathrm{(in/out)}} =
2\mathbf{T}^{\mathrm{(inc/scat)}} + \mathbf{I}$.
The incident/scattered formulation is much more commonly used; the
incoming/outgoing formulation gives simpler results for the transport
of momentum and angular momentum (that is, optical force and torque)
by the field. It should be note that for plane wave illumination, for which
the VSWF expansion is non-terminating, the incident/scattered formulation
gives a scattered wave expansion that converges over all space, while
the incoming/outgoing expansion, strictly used, would give an non-terminating,
non-convergent outgoing field expansion. For focussed beam illumination with
a finite VSWF expansion, the incoming/outgoing expansion directly gives
the total outgoing field that would be experimentally measured.
Since conversion from one formulation to the other is simple, either can
be readily used for calculation of fields, forces, scattering matrices,
or for orientation averaging.

In practice, the field expansions and the \textit{T}-matrix
are terminated at some $n = N_{\mathrm{max}}$. For the case of a
scatterer that is contained within a radius $r_0$, $N_{\mathrm{max}} \approx
kr_0$ is usually adequate, but $N_{\mathrm{max}} = kr_0 + 3 \sqrt[3]{kr_0}$ is
advisable if higher accuracy is needed~\cite{brock}. Although we assume in
this paper (as is usually the case) that the incident and scattered wave
expansions are terminated at the same $N_{\mathrm{max}}$ (giving a square
\textit{T}-matrix), this is not necessary. It should be noted that convergence
of the expansion of the incident field is not a necessary condition for the
\textit{T}-matrix method to be useful---indeed, for the most common
application, scattering of an incident plane wave, the incident field expansion
does not converge over all space. However, it does converge within the radius
$r_0$---which is the part of the field that can affect the
scattering particle---and therefore, the field expansions and the
\textit{T}-matrix can be truncated at a finite $N_{\mathrm{max}}$.

For the case of plane wave scattering, the plane wave expansion formula is
useful:
\begin{equation}
a_{nm} = 4\pi \mathrm{i}^n N_n \mathbf{C}_{nm}^\star \cdot \mathbf{E}_0,\;
b_{nm} = 4\pi \mathrm{i}^{n-1} N_n \mathbf{B}_{nm}^\star \cdot \mathbf{E}_0.
\end{equation}

The main case of interest for non-plane wave incident illumination is that of
focussed beams. A variety of methods can be used, such as plane wave
expansion~\cite{doicu1997ao}, the localised
approximation~\cite{gouesbet1995,gouesbet1996b,ren1998,polaert1998ao}, or the
point-matching method~\cite{nieminen_focussed}.

The only remaining requirement is that the \textit{T}-matrix be calculated.
This requires essentially a complete calculation of the scattering
properties of the scatterer. This is almost universally done using the extended
boundary condition method (EBCM), originally developed by
Waterman~\cite{waterman1971}, which is so strongly linked with the
\textit{T}-matrix method that the terms ``EBCM'' and ``\textit{T}-matrix
method'' are often used interchangeably. In the next section, we consider some
general principles involved in the calculation of the \textit{T}-matrix, and
show that an alternative method---column-by-column calculation using the
point-matching method (PMM)---is computationally feasible and simply
implemented for homogeneous isotropic particles devoid of symmetry.

Lastly, before we continue to consider calculation of \textit{T}-matrices
in more detail, we can note that while the incident and scattered fields are
usually expanded in terms of VSWFs, other sets of eigenfunctions,
such as cylindrical wavefunctions (for scatterers of infinite length in one
dimension), or a Floquet expansion (planar periodic scatterers), are more
appropriate for other geometries. There is no requirement that
the modes into which the incident and scattered fields are expanded be the same,
or even similar. In all of these cases, the \textit{T}-matrix method remains
applicable.

\section{Calculating the \textit{T}-matrix}

If the field expansions and \textit{T}-matrix are truncated at some
$N_{\mathrm{max}}$, there are
$N_T = 2 N_{\mathrm{max}} ( N_{\mathrm{max}} + 2 )$ expansion coefficients for
each of the incident and scattered fields, and the \textit{T}-matrix is
$N_T \times N_T$. Since $N_{\mathrm{max}}$ is proportional to the radius
enclosing the particle, $r_0$, the number of expansion coefficients is
proportional to $r_0^2$, and the number of elements in the \textit{T}-matrix is
proportional to $r_0^4$. This can be used to obtain an estimate of the scaling
of computational time for different methods of calculation.

\subsection{The extended boundary condition method}

In principle, any method of calculating scattering by the particle can be used
to calculate the \textit{T}-matrix. However, the method of choice is almost
universally the
EBCM~\cite{mishchenko_book,mishchenko1991,waterman1971,tsang1985}. In the EBCM,
the internal field within the particle is expanded in terms of regular VSWFs.
Therefore, the method is restricted to homogeneous and isotropic particles.
Rather than considering the coupling of the incident and scattered fields
directly, the coupling between the incident and internal (the $\mathbf{RgQ}$
matrix), and scattered and internal fields (the $\mathbf{Q}$ matrix) is
calculated, and the \textit{T}-matrix found from these
($\mathbf{T} = - \mathbf{RgQ} \mathbf{Q}^{-1}$). The $\mathbf{RgQ}$ and
$\mathbf{Q}$ matrices are the same size as the \textit{T}-matrix, with
$O(N_{\mathrm{max}}^4)$ elements. The elements of these matrices are found by
integrating over the surface of the scatterer, an operation requiring
$O(N_{\mathrm{max}}^2)$ time per element, so the calculation of the
$\mathbf{RgQ}$ and $\mathbf{Q}$ matrices is expected to require
$O(N_{\mathrm{max}}^6)$ computational time. The actual calculation of the
\textit{T}-matrix, if direct inversion is na\"{\i}vely used, takes
$O(N_{\mathrm{max}}^6)$ time. In practice, the calculation of the $\mathbf{RgQ}$
and $\mathbf{Q}$ matrices dominates the computational time~\cite{kahnert2001}.

From this, it can be seen that the EBCM can be expected to be very slow for
large particles. However, most applications of the EBCM have been for the
special case of scattering particles rotationally symmetric about the $z$ axis.
In this case, scattered modes $\mathbf{M}_{n'm'}^{(1)}$ and
$\mathbf{N}_{n'm'}^{(1)}$ only couple to incident modes $\mathbf{RgM}_{nm}$ and
$\mathbf{RgN}_{nm}$ if $m' = m$, greatly reducing the number of matrix elements
that need to be calculated, and the surface integral over the particle surface
reduces to a one-dimensional integral over $\theta$, since the azimuthal
integration over $\phi$ can be simply done analytically~\cite{tsang1985}. This
results in a great improvement in performance, and, in terms of computational
time, EBCM is clearly the method of choice for axisymmetric particles.
Numerical problems do occur when the scatterer is highly non-spherical. The
discrete sources method is designed to overcome these
problems~\cite{wriedt1998}. For the even more symmetric case of a spherical
scatterer, the scattered and incident modes only couple if $n' = n$ and
$m' = m$, the
\textit{T}-matrix becomes diagonal, and all of the integrals can be performed
analytically, and Mie's solution to scattering by a sphere~\cite{mie1908} is
simply obtained.

In a similar manner, scatterers with point-group rotational symmetry allow
significant improvement of the computational time required through exploitation
of the symmetry~\cite{kahnert2001,havemann2001,baran2001}.

Methods have also been developed to calculate \textit{T}-matrices for clusters
of particles and for layered particles~\cite{mishchenko_book}.

The efficiency of the
EBCM for the calculation of the \textit{T}-matrix is such that alternative
methods need only be considered if the EBCM is inapplicable (such as when the
particle in inhomogeneous or anisotropic), numerical difficulties are
encountered using the EBCM (such as for extremely non-spherical particles), or
if the scattering particle has no symmetries that can be used to optimise the
computation of the \textit{T}-matrix.

\subsection{Methods other than the EBCM}

Methods other than the EBCM can be used to calculate the \textit{T}-matrix. In
general, one would calculate the scattered field,  given a particular incident
field. The most direct way in which to use this to produce a \textit{T}-matrix
is to solve the scattering problem when the incident field is equal to a
single spherical mode---that is, a single VSWF such as
$\mathbf{E}_\mathrm{inc}(\mathrm{r}) = \mathbf{RgM}_{11}(k\mathrm{r})$,
$\mathbf{E}_\mathrm{inc}(\mathrm{r}) = \mathbf{RgN}_{11}(k\mathrm{r})$,
$\mathbf{E}_\mathrm{inc}(\mathrm{r}) = \mathbf{RgM}_{21}(k\mathrm{r})$,
etc, and repeat this for all VSWFs that need to be considered (up to
$n = N_{\mathrm{max}}$). The expansion coefficients for the scattered field can
be found in each case, if necessary, by using the orthogonal eigenfunction
transform (the generalised Fourier transform), and each scattering calculation
gives a single column of the \textit{T}-matrix.

Therefore, the calculation of a \textit{T}-matrix requires that
$2 N_{\mathrm{max}} ( N_{\mathrm{max}} + 2 )$ separate
scattering problems are solved. The provides a criterion for deciding whether it
is desirable to calculate a \textit{T}-matrix: if more than
$2 N_{\mathrm{max}} ( N_{\mathrm{max}} + 2 )$ scattering
calculations will be performed, then it is more efficient to calculate the
\textit{T}-matrix and use this for the repeated calculations than it is to use
the original scattering method repeatedly. Repeated calculations are expected if
orientation averaging is to be carried out, or if inhomogeneous illumination is
to be considered, such as, for example, scattering by focussed beams, where
there are generally 6 degrees of freedom, namely the three-dimensional position
of the scatterer within the beam, and the three-dimensional orientation of the
scatterer. Even if only a modest number of points are considered along each
degree of freedom, the total number of scattering calculations required rapidly
becomes very large, and even if the \textit{T}-matrix takes many hours to
calculate, the total time saved by doing so can make an otherwise
computationally infeasible problem tractable.

Volume methods are of interest, since they can readily be used for inhomogeneous
or anisotropic particles. The two most likely candidates are the
finite-difference time-domain method (FDTD)~\cite{mishchenko_book,yang2000} and
the discrete dipole approximation (DDA). In FDTD, the Maxwell equations are
discretised in space and time, and, beginning from a known initial state, the
electric and magnetic fields at each spatial grid point are calculated for
successive steps in time. The number of grid points required is
$O(N_{\mathrm{max}}^3)$ for three-dimensional scattering, and
$O(N_{\mathrm{max}})$ time steps required, so FDTD solutions scale as
$O(N_{\mathrm{max}}^4)$. Therefore, calculation of the \textit{T}-matrix using
FDTD should scale as $O(N_{\mathrm{max}}^6)$, which is the same scaling as the
EBCM. However, the grid required must be closely spaced compared to the
wavelength, and the space outside the scatterer must also be discretised, making
FDTD substantially slower than EBCM, especially for smaller particles. However,
FDTD is an extremely general technique, and has potential as a method for the
calculation of \textit{T}-matrices.

We should add that there is an additional consideration that makes FDTD
potentially
attractive as a method for calculating the \textit{T}-matrix: FDTD does not
assume that the incident wave is monochromatic. Consider the case when the
illumination is a brief pulse with a Gaussian envelope. The frequency spectrum
of the incident wave is Gaussian, and the scattering of a range of frequencies
can be found by taking the Fourier transform of the scattered
field~\cite{yee1991}. Even if we are not interested in other than monochromatic
illumination, we will frequently be interested in scattering by size
distributions of particles. Since varying the frequency for a particular
particle is equivalent to varying the size of the particle for a fixed incident
frequency, the \textit{T}-matrices for a range of particle sizes can be
calculated simultaneously.

The other major volume method for computational scattering, the discrete
dipole approximation (DDA),  also known as the coupled-dipole method, has been
recently applied to the calculation of the \textit{T}-matrix by
Mackowski~\cite{mackowski2002}, who obtained good results, with reasonable
computational efficiency using a moment method to solve the DDA system of
equations. DDA lacks the main disadvantages of FDTD, namely the need to
discretise space outside the particle, and the need to implement suitable
boundary conditions to prevent non-physical reflections from the boundary of the
computational domain. Mackowski's method scales as $O(N_{\mathrm{max}}^7)$ for
large $N_{\mathrm{max}}$. There is no need to discuss his
method in detail here, and the interested reader is
referred to his recent description of the method~\cite{mackowski2002}.

Finally, we consider the point-matching method. Like the \textit{T}-matrix
method and the EBCM, the point-matching method involves expansion of fields
in terms of VSWFs. In the point-matching method, the internal field within the
scatterer and the scattered field are expanded as series of VSWFs, and the
incident, internal, and scattered fields are matched at points on the particle
surface, using the usual boundary condition of continuity of tangential
components of the electric and magnetic fields. This gives a system of equations
from which the unknown expansion coefficients of the internal and scattered
fields can be found. Typically, enough points are used for matching the fields
so as to give an overdetermined system of equations, which is then solved in a
least-squares sense. Solving the system of $O(N_{\mathrm{max}}^2)$ unknowns
using direct matrix inversion can be expected to be an $O(N_{\mathrm{max}}^6)$
problem,  with the result that the total computational time is
$O(N_{\mathrm{max}}^8)$. In practice, faster methods can be used, and our
results indicate a performance of about $O(N_{\mathrm{max}}^7)$ for our
implementation.

The point-matching method is an attractive candidate since a \textit{T}-matrix
implementation will generally include routines to calculate VSWFs, making the
implementation of a point-matching \textit{T}-matrix calculator simple. The only
further requirement is a routine for solving overdetermined linear systems.
Since the scattered field is calculated in terms of VSWFs, the conversion of the
results of a single PMM calculation to a \textit{T}-matrix column is trivial.

Naturally, multiple expansion methods (the generalised multipole technique, or
the multiple multipole method) can  be used. Since multipole methods exist for
anisotropic media~\cite{piller}, the method can be used for anisotropic
scatterers.

Our implementation of the point-matching method, and its performance,
is discussed in the next section.

\section{Point-matching method}

Our implementation of the PMM \textit{T}-matrix calculation uses an
incoming/outgoing field expansion (equations (\ref{incoming_expansion}) and
(\ref{outgoing_expansion})), rather than the usual incident/scattered wave
expansion (equations (\ref{regular_expansion}) and (\ref{outgoing_expansion})),
and the internal field is expanded in terms of regular VSWFs:
\begin{eqnarray}
\mathbf{E}_\mathrm{inc}(\mathrm{r}) & = &
\sum_{n=1}^{N_{\mathrm{max}}} \sum_{m = -n}^n
a_{nm} \mathbf{M}_{nm}^{(2)}(k\mathrm{r}) +
b_{nm} \mathbf{N}_{nm}^{(2)}(k\mathrm{r}),
\label{expansion_e_inc}\\
\mathbf{E}_\mathrm{scat}(\mathrm{r}) & = &
\sum_{n=1}^{N_{\mathrm{max}}} \sum_{m = -n}^n
p_{nm} \mathbf{M}_{nm}^{(1)}(k\mathrm{r}) +
q_{nm} \mathbf{N}_{nm}^{(1)}(k\mathrm{r}),
\label{expansion_e_scat}\\
\mathbf{E}_\mathrm{int}(\mathrm{r}) & = &
\sum_{n=1}^{N_{\mathrm{max}}} \sum_{m = -n}^n
c_{nm} \mathbf{RgM}_{nm}(k\mathrm{r}) + d_{nm} \mathbf{RgN}_{nm}(k\mathrm{r}).
\label{expansion_e_int}
\end{eqnarray}
We use this particular expansion since we are interested in calculating optical
forces and torques within optical traps~\cite{nieminen2001jqsrt,nieminen2001cpc}
and this results in simpler expressions for these quantities.

We considered a single scatterer, centred on the origin,
contained entirely within a radius $r_0$, and with a surface specified by a
function of angle:
\begin{equation}
r = r(\theta,\phi)
\end{equation}

The boundary conditions---matching the tangential fields on the surface of the
scatterer---are
\begin{eqnarray}
\hat{\mathbf{n}} \times ( \mathbf{E}_\mathrm{inc}(\mathrm{r}) +
\mathbf{E}_\mathrm{scat}(\mathrm{r}) ) & = &
\hat{\mathbf{n}} \times \mathbf{E}_\mathrm{int}(\mathrm{r}),\\
\hat{\mathbf{n}} \times ( \mathbf{H}_\mathrm{inc}(\mathrm{r}) +
\mathbf{H}_\mathrm{scat}(\mathrm{r}) ) & = &
\hat{\mathbf{n}} \times \mathbf{H}_\mathrm{int}(\mathrm{r}),
\end{eqnarray}
where $\hat{\mathbf{n}}$ is a unit vector normal to the surface of the particle.

The magnetic fields are given by expansions similar to those for the electric
fields:
\begin{eqnarray}
\mathbf{H}_\mathrm{inc}(\mathrm{r}) & = & \frac{1}{k_{\mathrm{medium}}}
\sum_{n=1}^{N_{\mathrm{max}}} \sum_{m = -n}^n
a_{nm} \mathbf{N}_{nm}^{(2)}(k\mathrm{r}) +
b_{nm} \mathbf{M}_{nm}^{(2)}(k\mathrm{r}),\\
\mathbf{H}_\mathrm{scat}(\mathrm{r}) & = & \frac{1}{k_{\mathrm{medium}}}
\sum_{n=1}^{N_{\mathrm{max}}} \sum_{m = -n}^n
p_{nm} \mathbf{N}_{nm}^{(1)}(k\mathrm{r}) +
q_{nm} \mathbf{M}_{nm}^{(1)}(k\mathrm{r}),\\
\mathbf{H}_\mathrm{int}(\mathrm{r}) & = & \frac{1}{k_{\mathrm{particle}}}
\sum_{n=1}^{N_{\mathrm{max}}} \sum_{m = -n}^n
c_{nm} \mathbf{RgN}_{nm}(k\mathrm{r}) + d_{nm} \mathbf{RgM}_{nm}(k\mathrm{r}).
\end{eqnarray}
where $k_{\mathrm{medium}}$ and $k_{\mathrm{particle}}$
are the wavenumbers of the field in the surrounding medium and inside the
particle, respectively.

There are $4 N_{\mathrm{max}} ( N_{\mathrm{max}} + 2 )$ unknown variables---the
expansion coefficients $c_{nm}$, $d_{nm}$, $p_{nm}$,  and $q_{nm}$. Since the
fields are vector fields, each point gives multiple equations---four
independent equations per point. We generate a grid of
$2 N_{\mathrm{max}} ( N_{\mathrm{max}} + 2 )$ points with equal angular
spacings in each of the $\theta$ and $\phi$ directions,  giving
$8 N_{\mathrm{max}} ( N_{\mathrm{max}} + 2 )$ independent equations. Equal
angle spaced points are used for simplicity, although points uniformly
distributed about a sphere would be better~\cite{wriedt_gainesville}.

The values of the VSWFs at these points on the particle surface are calculated,
and used in the column-by-column calculation of the \textit{T}-matrix.

The computation time (which is independent of the particle shape, depending only
on the containing radius $r_0$) is shown in table~\ref{computation_time}. The
calculations were carried out in MATLAB~\cite{matlab} on a 1.5\,GHz PC. The
times taken are reasonable in comparison to computation times for
EBCM for particles with no symmetry~\cite{kahnert2001}.

\begin{table}
\begin{center}
\begin{tabular}{cr@{.}lr@{.}l}
\hline
$N_{\mathrm{max}}$ & \multicolumn{2}{c}{$kr_{0\mathrm{max}}$}
& \multicolumn{2}{c}{Time}\\
\hline
 1 & 0&033 & 0&041\,s \\
 2 & 0&21  & 0&16\,s \\
 3 & 0&55  & 0&85\,s \\
 4 & 1&00  & 7&00\,s \\
 5 & 1&54  & 30&3\,s \\
 6 & 2&14  & 1&86\,min \\
 7 & 2&78  & 4&95\,min \\
 8 & 3&46  & 12&2\,min \\
 9 & 4&17  & 26&8\,min \\
10 & 4&90  & 56&3\,min \\
11 & 5&66  & 1&91\,h \\
12 & 6&42  & 3&53\,h \\
13 & 7&21  & 6&35\,h \\
\hline\\
\end{tabular}
\end{center}
\caption{Computation times for calculating \textit{T}-matrices. The
calculations were carried out in MATLAB on a 1.5\,GHz PC. The maximum size
parameter $kr_0$ for which the truncation is expected to always be
well-convergent is shown. Reasonable convergence can also be expected for
size parameters $kr_0 \approx N_{\mathrm{max}}$}
\label{computation_time}
\end{table}

\begin{figure}[htb]
\centerline{\includegraphics[width=0.6\columnwidth]{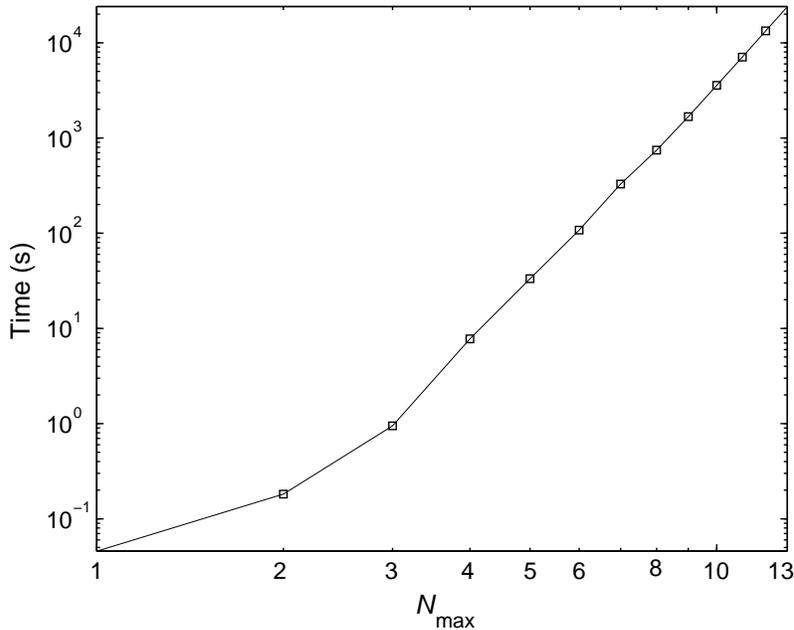}}
\caption{Computation times for calculating \textit{T}-matrices. The
calculations were carried out in MATLAB on a 1.5\,GHz PC. The time taken
scales as $O(N_{\mathrm{max}}^{6.8})$ for $N_{\mathrm{max}} > 2$.}
\label{computation_time_graph}
\end{figure}

\begin{figure}[htb]
\centerline{\includegraphics[width=0.6\columnwidth]{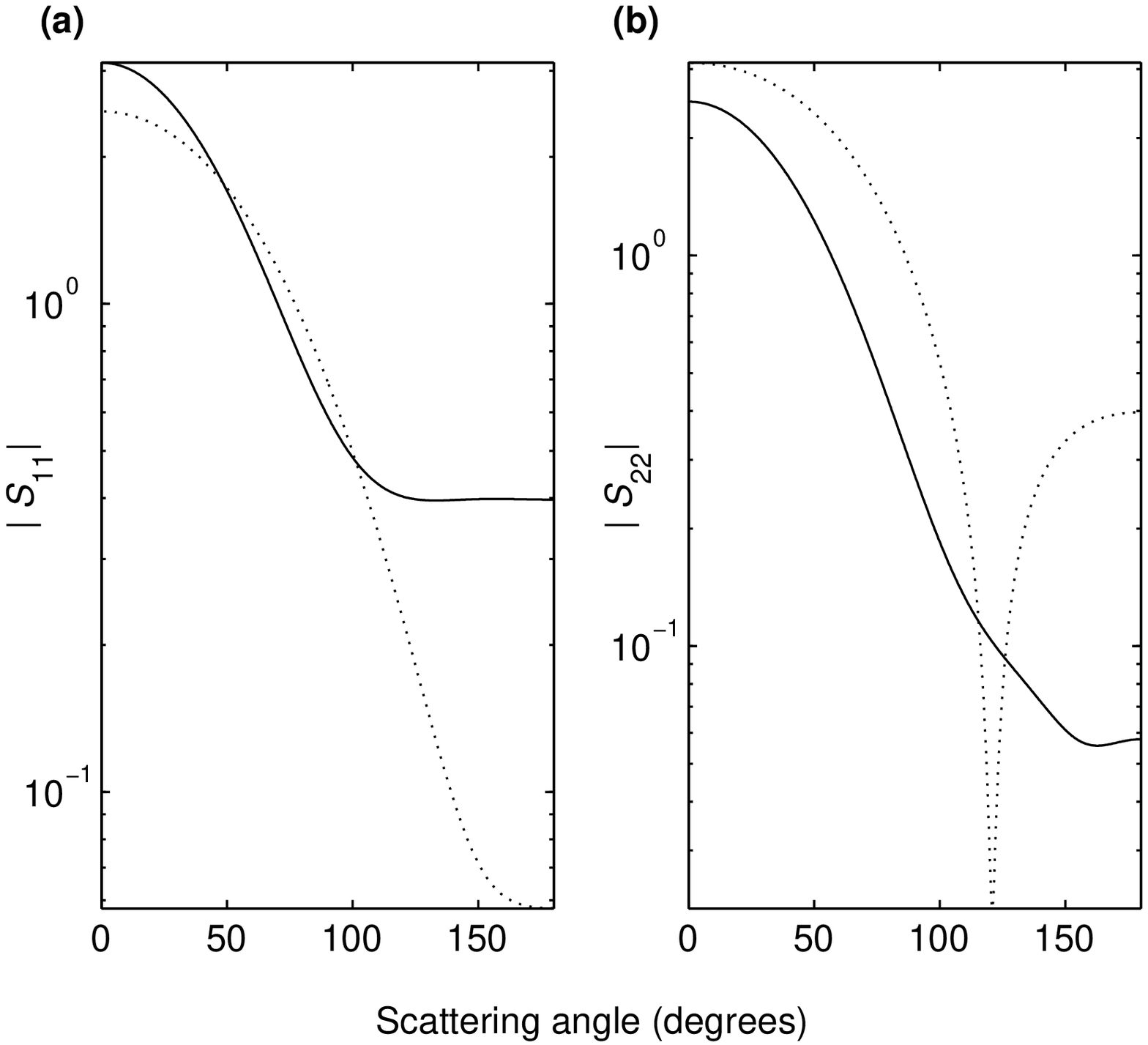}}
\caption{(a) $S_{11}$ and (b) $S_{22}$ scattering matrix elements for an
ellipsoidal particle with axes of length $a = 1\lambda$, $b = 0.2\lambda$, and
$c = 0.5\lambda$, and relative refractive index $m = 1.5$. The incident
illumination is directed along the $c$ axis of the ellipsoid. The solid
line shows scattering in the $ac$ plane (containing the largest area
cross-section of the ellipsoid), and the dotted line shows scattering in the
$bc$ plane (the smallest cross-sectional area).}
\label{scattering_matrix}
\end{figure}

Results of a sample calculation are shown in figure~\ref{scattering_matrix},
where the diagonal scattering matrix elements are shown, calculated using the
PMM \textit{T}-matrix. The scattering matrix elements $S_{11}$ and $S_{22}$ are
shown for scattering in two different planes; the effect of non-axisymmetry is
evident.

The accuracy and validity of the PMM-calculated \textit{T}-matrix will be
essentially the same as the accuracy and validity of the point-matching
algorithm used in the calculation. Thus, a detailed analysis of our
simple proof-of-principle implementation serves little purpose. It is obviously
useful to use the best, sufficiently fast, point-matching code available. In
view of the mathematical similarity between the \textit{T}-matrix method and the
point-matching method, it should be a simple task to adapt any PMM code to the
task of generating \textit{T}-matrix columns.

\section{Conclusions}

The point-matching method is suitable for the calculation of the
\textit{T}-matrix for particles with no symmetry, provided that the particles
are not too large. The method has the advantage of being extremely simple to
implement within a general \textit{T}-matrix package, since most of the required
routines will be shared with the existing \textit{T}-matrix code.
This results from the mathematical formalisms of the \textit{T}-matrix
method and the point-matching method being essentially the same. Any
point-matching algorithm can be used, with multiple expansion origins,
automatic convergence checks, and so on.
Since the PMM uses the same field expansions as the EBCM, the same
numerical difficulties are to be expected for scatterers with large aspect
ratios; in such cases,  multiple expansion origin algorithms will be
necessary.
The accuracy of the PMM
\textit{T}-matrix will be the same as the PMM which is used to calculate it.
Naturally, the usual conditions of applicability of the PMM, such as the
validity of the Rayleigh hypothesis, need to be considered.

The PMM explicitly depends on the Rayleigh hypothesis---the
assumption that the fields can be represented by the expansions
(\ref{expansion_e_inc})---(\ref{expansion_e_int}) over all space
rather than just outside and inside spherical surfaces circumscribing and
inscribing the surface of the scatterer.
The validity of this assumption for arbitrary scatterers is unknown.
However, the use of an overdetermined system of equation may well extend
the method somewhat beyond the strict range of applicability of the
Rayleigh hypothesis by providing a least squares approximation of the
fields between the circumscribing and inscribing surfaces where the
VSWF expansions might be non-convergent.
One advantage of relying on the Rayleigh hypothesis is that the
fields are given everywhere, including the fields internal to the
scatterer (a \textit{T}$^{(\mathrm{int})}$-matrix can be used to
relate the internal and incident fields). This applies generally
to methods that make use of the Rayleigh hypothesis, such as
the generalised separation of variables method~\cite{rother1998}.
In contrast to this, the EBCM, which avoids the Rayleigh hypothesis,
gives the tangential surface fields on the surface of the scatterer, rather
than the internal fields.

The point-matching method lacks the generality of DDA and FDTD.
In this respect, the recent discrete dipole moment method
\textit{T}-matrix calculations by Mackowski~\cite{mackowski2002} are
particularly promising.

Lastly, we note again that FDTD may prove to be a useful method for
\textit{T}-matrix calculation since it can be used to calculate
\textit{T}-matrices simultaneously for a range of particle sizes.

\end{document}